\documentclass[12pt,aps,prl,floatfix,tightenlines]{revtex4}
\pdfoutput=1
\usepackage[pdftex]{hyperref}
\usepackage{amsmath}
\usepackage{latexsym}
\usepackage{graphicx}
\usepackage{color}

\begin{document}

\title{``Quantum-optical coherence tomography'' with classical light}
\author{J. Lavoie, R. Kaltenbaek, and K. J. Resch}
\email{kresch@iqc.ca}
\affiliation{Institute for Quantum Computing and Department of Physics \&
Astronomy, University of Waterloo, Waterloo, Canada, N2L 3G1}

\begin{abstract} Quantum-optical coherence tomography (Q-OCT) is an
interferometric technique for axial imaging offering several
advantages over conventional methods.  Chirped-pulse interferometry
(CPI) was recently demonstrated to exhibit all of the benefits of
the quantum interferometer upon which Q-OCT is based.  Here we use
CPI to measure axial interferograms to profile a sample accruing the
important benefits of Q-OCT, including automatic dispersion
cancellation, but with 10 million times higher signal.  Our
technique solves the artifact problem in Q-OCT and highlights the
power of classical correlation in optical imaging.
\end{abstract}
\maketitle

\section{Introduction}
Optical coherence tomography (OCT) \cite{fujimoto95} is a
non-invasive imaging technique using low-coherence interferometry to
produce depth profiles of a sample.  OCT has found many biomedical
applications; prominent examples include the diagnosis of ocular
diseases or detection of early-stage cancer \cite{fercher03}. Axial
resolution in OCT is ultimately limited by the coherence length of
the light source and can be less than 1~$\mu$m for very broadband
sources \cite{drexler04}.  This resolution is hindered by material
dispersion which both broadens the features in the
interferograms and reduces the contrast.

Exciting developments in quantum interferometry led to the proposal
and demonstration of quantum-optical coherence tomography (Q-OCT)
\cite{abouraddy02,nasr03}. This technique replaces white-light
interferometry (WLI) with Hong-Ou-Mandel (HOM) interferometry
\cite{hong87} utilizing frequency-entangled photon pairs.  This
device automatically cancels all even orders of dispersion in the
resulting interferogram, including group-velocity dispersion, the most
significant contribution \cite{steinberg92b}. Dispersion
cancellation in HOM interference is ``blind'', requiring no \emph{a priori}
knowledge of the material properties, in contrast with dispersion
compensation methods (see \cite{fercher03} and references therein).
In addition to dispersion cancellation, the HOM interferometer is
phase insensitive, has better resolution than WLI with the same
bandwidth, and the interference visibility is insensitive to
unbalanced loss.  Unfortunately, this technique is based on entangled
photon pairs and the costs, in terms of speed and expense, have limited its
widespread adoption.  Other
techniques for blind dispersion compensation without entanglement
have been proposed \cite{erkmen06,banaszek07} or demonstrated
\cite{resch07b}, but they require unavailable technology
\cite{erkmen06} or significant numerical post-processing
\cite{banaszek07,resch07b} and do not have the other advantageous
properties of Q-OCT.  We have recently
demonstrated a completely classical technique, based on
oppositely-chirped laser pulses, for producing an interferogram with
\emph{all} the advantages of HOM interference with vastly higher
signal~\cite{kaltenbaek08}.

In the present work, we use chirped-pulse interferometry (CPI) for axial
profiling a sample with two optical interfaces. Q-OCT interferograms have
been shown to contain artifacts \cite{abouraddy02}, signals that do not
correspond to real features of the sample.  Due to the strong analogy,
these artifacts also appear in CPI-based axial imaging. We experimentally
demonstrate a straightforward method for controlling these artifacts.
Although it is possible to control these artifacts in Q-OCT as
well~\cite{abouraddy02}, it is technically challenging and has not been
demonstrated.

\begin{figure}
\begin{center}
\includegraphics[width=1 \columnwidth]{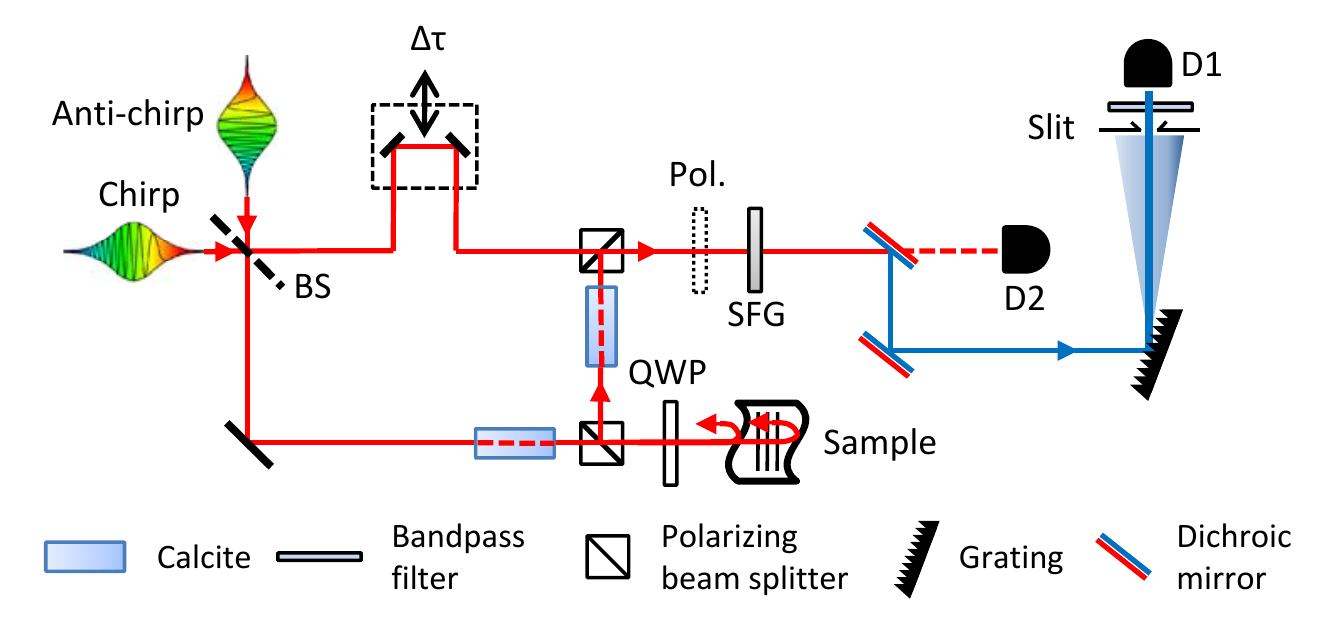}
\caption{Experimental setup for axial profiling with chirped-pulse
interferometry.  Pairs of oppositely-chirped laser pulses with
horizontal polarization are combined at a 50/50 beam splitter (BS).
The light from one BS output reflects from a sample; the light from
the other undergoes a spatial delay. In the sample arm, two passes
through the quarter-wave plate (QWP) rotate the polarization to
vertical. This allows spatial recombination of the two beams at a
polarizing beam splitter. Both beams are focussed onto a 0.5~mm
thick BBO crystal phase-matched for type-II sum-frequency generation
(SFG). Dichroic mirrors separate the fundamental from the SFG light.
A grating and slit are used to filter a narrow band (0.46~nm FWHM)
of SFG light before the light is detected by an amplified Si
photodetector (D1).  An alternate configuration, where a 45$^\circ$
polarizer is inserted before the nonlinear crystal and the
fundamental light is directly detected with a photodiode (D2),
allows the observation of white-light fringes and a direct
comparison with CPI. A pair of calcite blocks can be inserted to
compare the effects of material dispersion on the interferograms.
\label{setup}}
\end{center}
\end{figure}

\section{Experimental setup and methods}
Our interferometer is shown in Fig.~1 and described in the caption.  It
relies on pairs of
oppositely-chirped laser pulses that have been
stretched to several hundred times their initial, transform-limited,
pulse duration. In this large-chirp limit, at any given time the two
frequency-anticorrelated pulses have frequencies $\omega_0+\Omega$ and
$\omega_0-\Omega$, respectively. Here, $\omega_0$ is the average of the
instantaneous frequencies of these pulses.
If the chirped pulses are coincident at the input beam splitter,
$\omega_0$ is equal to the centre frequency of the laser, but it can
be tuned by changing the relative delay between the pulses. We refer
to $\omega_0$ as the \emph{operating frequency} to distinguish it
from the centre frequency of the laser.  Following the theoretical
framework from \cite{abouraddy02}, we assume that the effect of
the sample is modelled by a linear transfer function, $H(\Omega)$. The
reference arm contains an adjustable path delay, $\Delta\tau$. After
propagation in each arm, the beams undergo sum-frequency generation
(SFG) in a nonlinear medium. We detect SFG light in a very narrow
frequency band near $2\omega_0$ ensuring that the output signal is
almost exclusively due to cross-correlations between the chirped and
anti-chirped pulses. Under these conditions, the signal integrated over all
frequencies in the chirped pulses and measured by a
square-law detector, $S(\Delta\tau)$, is
\begin{eqnarray}
S(\Delta\tau) &\propto& \int d\Omega I(\Omega)I(-\Omega)|H(\Omega)|^2
- \textrm{Re} \left[ \int d\Omega
I(\Omega)I(-\Omega)H(\Omega) H^*(-\Omega) e^{-2i\Omega \Delta\tau}
\right],
\end{eqnarray}
\noindent where $I(\Omega)$ is the intensity spectrum of both laser
pulses. The CPI
signal is identical to that in Q-OCT when $I(\Omega)I(-\Omega)$ is
equal to the spectrum of the entangled photons (see \cite{abouraddy02}, Eqns.~(6)-(8)).

A mode-locked Ti:Sapphire laser (center wavelength 790~nm, average power 2.7~W,
repetition rate 80 MHz) was used to create a pair of horizontally-polarized
beams of oppositely-chirped pulses using a grating-based stretcher
and compressor \cite{pessot87}.  These stretched the initial pulses
from 100~fs to 54~ps (48~ps) with 11~nm (10~nm) bandwidth for the
chirped (anti-chirped) pulses.  Note that the difference in the
pulse duration is due to slightly different bandwidths, not different chirp
rates. Details on the stretcher and compressor can be found in
\cite{kaltenbaek08}.

\section{Results and discussions}
We used a borosilicate microscope coverglass as the sample. The CPI
(WLI) data was taken by recording the signal of detector D1 (D2)
over the delay $\Delta\tau$. In each scan data was accumulated for
0.5~s over a range of 0.5~mm. The data without and with the calcite
blocks are shown in Fig.~2(a) and Fig.~2(b), respectively, where the
upper (lower) plots are CPI (WLI) scans. Without additional
dispersion (Fig.~2(a)), the CPI interference dips have widths of
20.1$\pm$0.3~$\mu$m and 20.6$\pm$0.4~$\mu$m FWHM and corresponding
visibilities of 39.0\% and 40.9\% for the front and the back surface
of the coverglass, respectively.  The WLI patterns have widths
30.4$\pm$0.3 $\mu$m and 29.9$\pm$0.3 $\mu$m FWHM and corresponding
visibilities of 12.9\% and 14.0\%. The errors in these measurements
are mainly due to positioning uncertainty of the motor. Even without
the addition of calcite blocks, we see that the CPI signal has
enhanced resolution by a factor of 1.5 compared to WLI. This
enhancement factor is slightly larger than the theoretically
expected value of $\sqrt{2}$ (assuming Gaussian spectra), due to
uncompensated dispersion caused by the additional PBS in the sample
arm.
\begin{figure*}
\begin{center}
\includegraphics[width=1 \columnwidth]{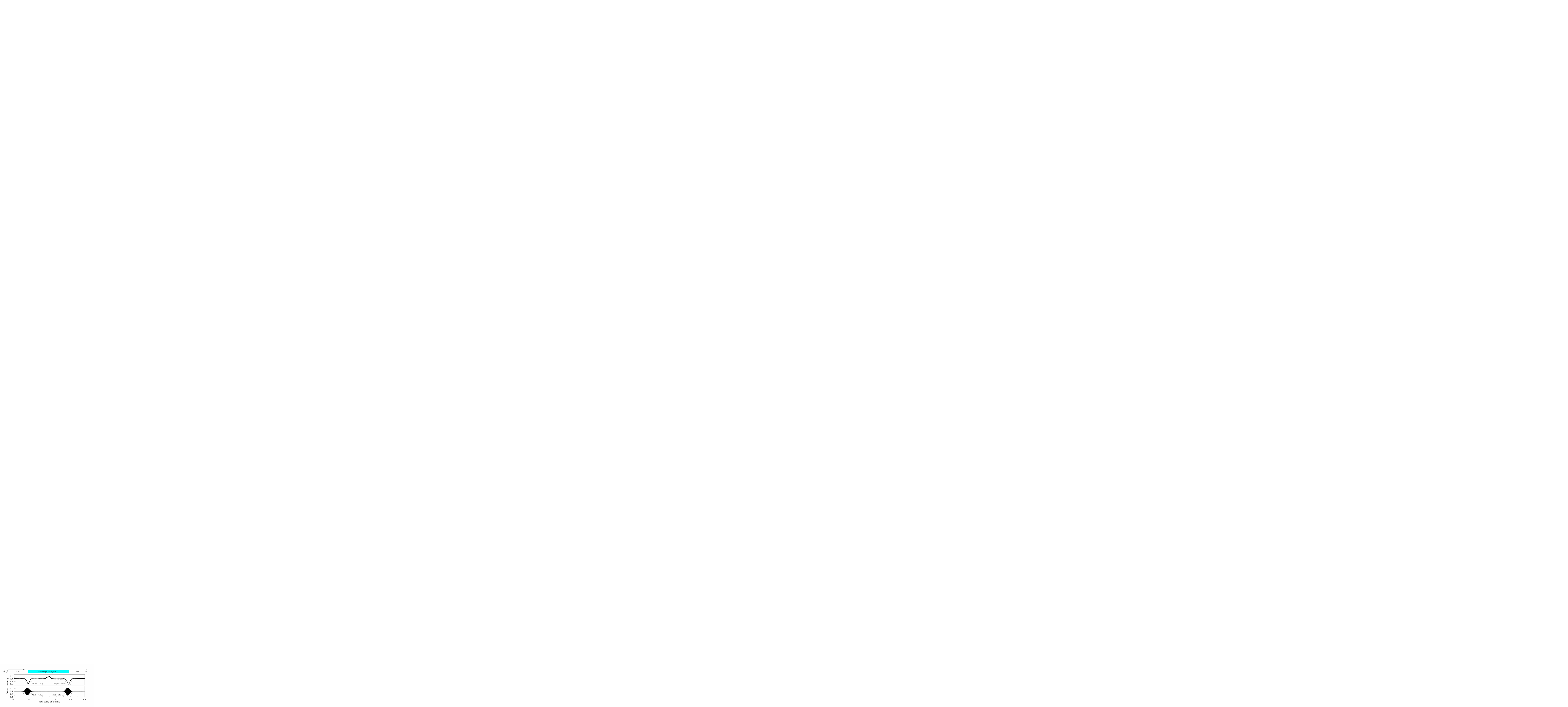}
\includegraphics[width=1 \columnwidth]{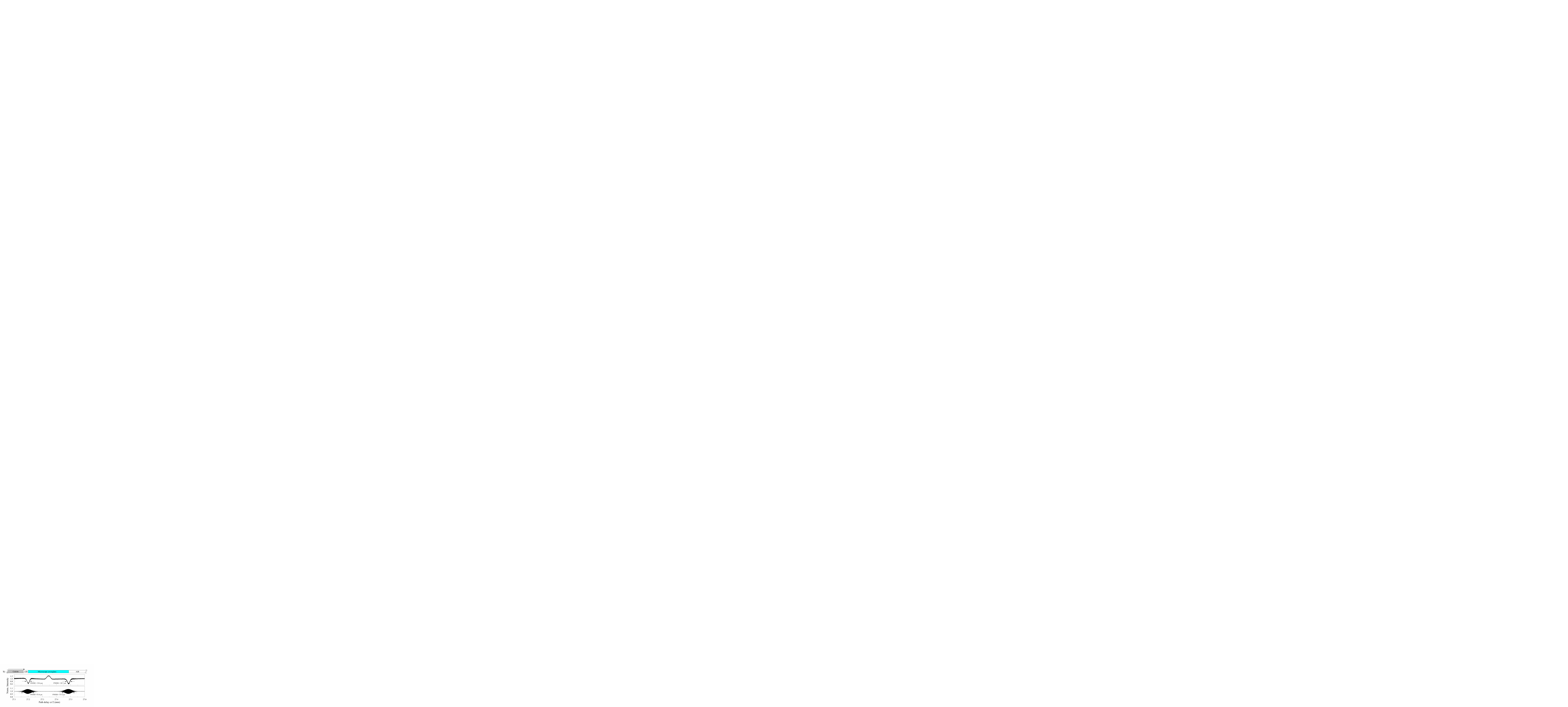}
\caption{Axial scans of a microscope coverglass using chirped-pulse and
white-light interference. Light enters from the left and is reflected from
either the front or the back surface of the sample, as indicated at the top of
the figure. The normalized detector signal is plotted as a
function of path delay. Each data set shows interference features
corresponding to the front and back surface reflections of the
sample.  The CPI (top) and the WLI (bottom) were taken a) without
and b) with calcite blocks.  The CPI signal resolution, as measured
by the width of the interference feature, is unaffected by
the dispersion whereas the WLI is broadened by 74\%.  As in Q-OCT,
CPI shows an artifact between the two real signals. \label{scans}}
\end{center}
\end{figure*}

The average optical power in the WLI scans in Figs.~2(a)~\&~2(b) was
\mbox{$\sim$10~mW}. In the CPI scans, the normalized intensity of
1.0 corresponds to 0.3~$\mu$W of measured optical power. This
corresponds to approximately $10^{12}$~photons/s. It is difficult to
directly compare this rate with the performance of Q-OCT
demonstrated in \cite{nasr03} since only normalized rates are
presented; however, the highest measured coincidence-counting rate
from an entangled photon source to date is $\sim10^6$ photons/s
\cite{altepeter05}. Given a total sample reflectivity of $\sim10\%$
our measured power is 7 orders of magnitude larger than what could
be achieved in Q-OCT with the best available technology.

\begin{figure}
\begin{center}
\includegraphics[width=1 \columnwidth]{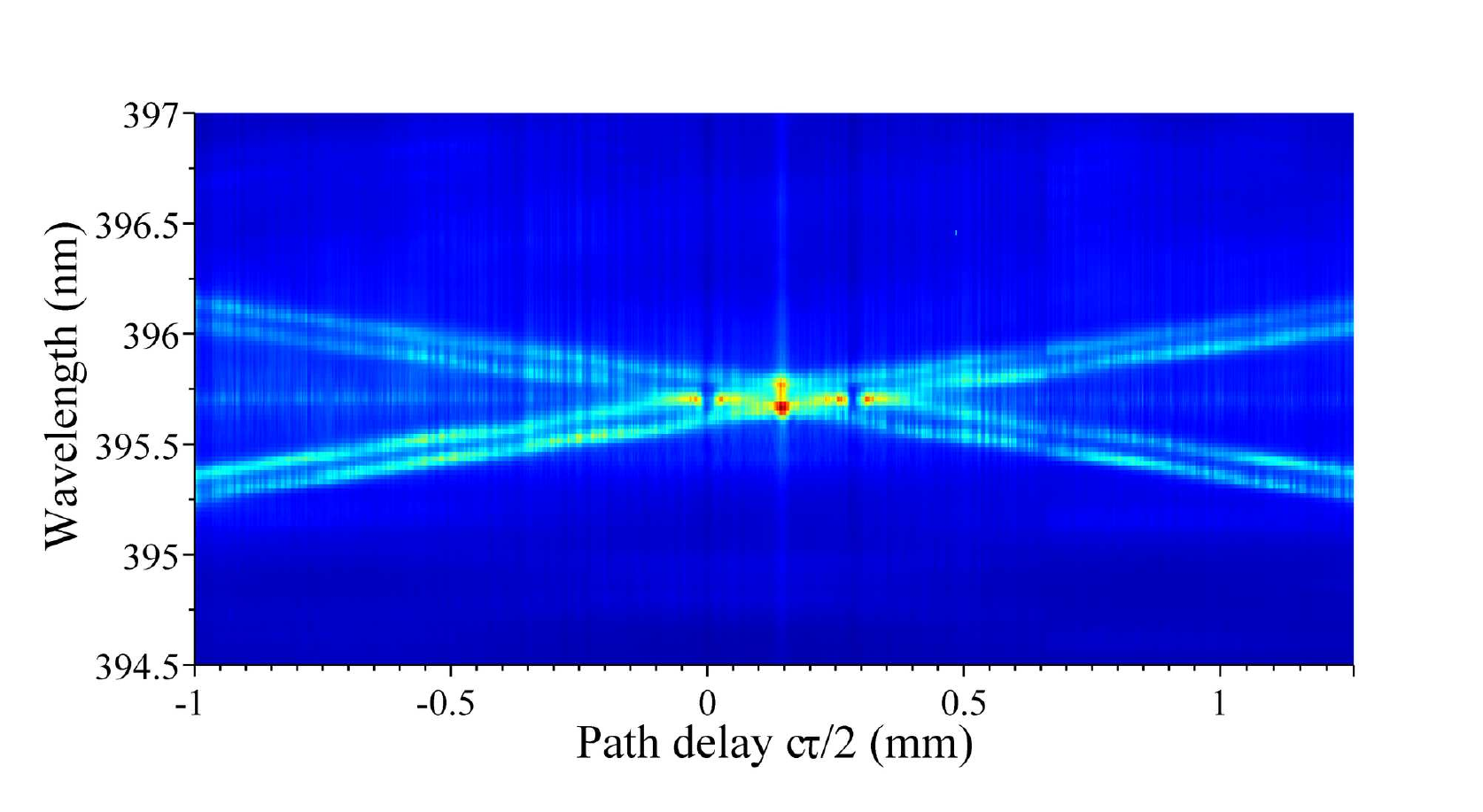}
\caption{False-color representation of the SFG spectrum vs path delay.
Two pairs of narrow lines originate from SFG of the
oppositely-chirped laser pulses with different time delays. When the
time delay through the reference arm coincides with the delay
through the sample arm from one of the two interfaces, an
interference dip occurs.  The other pair of crossings between the
real interference dips gives rise to the artifact.\label{falsecolor}}
\end{center}
\end{figure}

The path delay between the two CPI interference dips is
286.1$\pm$0.4~$\mu$m. The operating frequency was measured by taking
the spectrum of the light after the first beam splitter; it is the
frequency at which the chirped and antichirped pulses interfere. The
observed delay can be converted to the thickness of the coverslip by
dividing by the group index, $n_g(\lambda)=n(\lambda)-\lambda \left. \frac{dn}{d\lambda}
\right|_\lambda$, of borosilicate
glass at the operating wavelength $790.8\pm0.3$~nm, $n_g=1.53482$
\cite{borosellmeier}.  The optical measurement of the coverglass
thickness is $186.4\pm0.3$~$\mu$m which is in good agreement with a
direct measurement, using a micrometer, yielding
$186.4\pm0.8$~$\mu$m.

To investigate the effects of material dispersion on the
interference, we added a pair of calcite beam displacers into the
setup. The sum of their lengths is 80.58$\pm$0.01~mm and the light
propagates through them with ordinary polarization. The widths of
the CPI dips are unchanged at 19.4$\pm$0.6~$\mu$m and
20.7$\pm$0.4~$\mu$m FWHM; the WLI envelopes are significantly
broadened by 74\% to 53.0$\pm$0.3 $\mu$m and 53.4$\pm$0.3 $\mu$m.
Under these conditions, CPI has a factor of 2.65 better resolution
than WLI.

Both CPI and Q-OCT signals contain artifacts, additional features in
the interferograms that do not correspond to real interfaces.  These
can be seen in the data in Fig.~2 in between the interference dips.
To illuminate the origin of these features in CPI we measured the
full SFG spectrum as a function of delay using a high-resolution
spectrometer (ACTON SP-2758).  The results are shown in Fig.~3. On
this scale, the only features visible are due to cross-correlations;
the autocorrelations form a weak, broadband background.  When the
paths are unbalanced, the signal contains two doublets of narrow
spectral lines.  One of these doublets is due to the chirped pulse
traversing the sample arm and the anti-chirped pulse traversing the
reference. The two peaks of the doublet are separated in frequency
due to the difference in optical delay reflecting from the front and
back surface of the sample. When the chirped pulse is reflected from
the front surface the frequency of the cross-correlation will be
slightly higher than upon reflection from the back surface. The
other doublet can be understood by swapping the roles of the chirped
and anti-chirped pulses.

Changing the path delay changes the spacing
between the two doublets. An interference dip occurs when the delay
in the reference path is equal to a delay in the sample path. The
interference results from two different processes (chirp in one of
the two arms and anti-chirp in the other) each producing light at
the same frequency, but out of phase. In Fig.~3 the four lines,
formed by the two doublets for varying path delay, cross at four
distinct points. Two of these points occur at the same wavelength
but at different path delays; they correspond to the dips in the CPI
scans indicating real features of the sample. The other two crossing
points occur at the same path delay but different wavelengths; these
give rise to the artifacts.

\begin{figure}
\begin{center}
\includegraphics[width=1 \columnwidth]{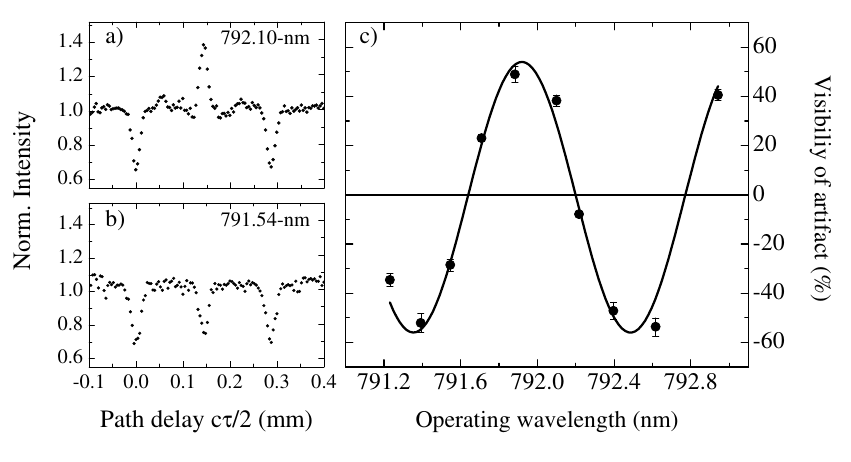}
\caption{Controlling the phase of the artifact.  CPI interferograms
of the sample taken at an operating wavelength of a) 792.10~nm and
b) 791.54~nm clearly shows the dependence of the phase of the
artifact interference on the operating wavelength. c) Visibility of
the artifact versus operating wavelength. Positive (negative)
visibility corresponds to constructive (destructive) interference.
The measured period of oscillation of (1.13$\pm$0.02)~$n$m is in
good agreement with the theoretical expectation.\label{fringes}}
\end{center}
\end{figure}

As in Q-OCT, the interference giving rise to the artifacts can be
constructive or destructive whereas for real features it is always
destructive. If the interference is constructive, as in Fig.~2, the
artifact is easy to identify.  If instead it is destructive,
artifacts can easily be confused with real features. In Q-OCT it was
predicted that this could be adjusted by changing the sum-frequency
of the entangled photon pair~\cite{abouraddy02}. In practice,
however, this is difficult as most UV narrow-band pump sources for
SPDC are not tunable. CPI has the intrinsic advantage that the
operating frequency is tunable by changing the relative delay
between the chirped and anti-chirped pulses at the input beam
splitter. In Figs.~4(a) and 4(b), we show two examples of CPI
interferograms taken at different operating wavelengths 792.10~nm
and 791.54~nm illustrating constructive and destructive interference
in the artifact, respectively. The operating frequency is half the
SFG frequency measured near zero delay.  In these scans the path
delay was not varied continuously but in discrete steps, accounting
for fewer data points.

We employ the model transfer function for the coverslip,
$H(\Omega)=r_1+r_2 e^{i2k(\omega_0+\Omega)d}$, where $r_1$ ($r_2$)
is the reflection amplitude from the front (back) surface,
$k(\omega)$ is the wavevector in the glass, and $d$ is the
thickness. Inserting this expression into Eq.~(1), one finds that the
term describing the artifact is modulated by $\cos 2 k(\omega_0) d$. If
the operating frequency changes from $\omega_0$ to $\omega_0+\delta$, then
$k(\omega_0+\delta)\approx k(\omega_0)+\alpha \delta$ and the expected
change in wavelength required to flip the sign of the artifact is
$\Delta \lambda \approx \pi^2 c/(\omega^2_0\alpha d)$.

Figure~4(c) shows the visibility of the artifact as a
function of the operating wavelength.  Visibility is defined as $(I_C-I_S)/I_S$, where
$I_C$ and $I_S$ are the intensities at the centre of the
dip and at the shoulder, respectively. A fit to this data yields a
period of (1.13$\pm$0.02)~nm in good agreement with the theoretical
prediction of 1.09~nm. Changing the operating wavelength in this
straightforward way allows identification and removal of artifacts
from axial scans.

There is a difference between Q-OCT and CPI. Even in the absence of
dispersion, the resolution in Q-OCT is a factor of $2$ better than
in WLI~\cite{abouraddy02} while, assuming Gaussian spectra, CPI has
a factor of $\sqrt{2}$ better resolution than WLI. The difference
originates from the effective bandwidths used in the comparison.
Assuming a source of entangled photons with spectrum $S(\Omega)$
(the modulus squared of $\zeta(\Omega)$ in Eq.~(5) in
\cite{abouraddy02}), the effective bandwidths for HOM
interference (see \cite{abouraddy02}, Eq.~(8)) and WLI (see
\cite{abouraddy02}, Eq.~(4)) are the same because their spectra,
$S(\Omega)$, are identical \cite{widthfootnote}. For chirped pulses
with spectra $I(\Omega)$, the bandwidth for CPI is determined by
$I(\Omega)I(-\Omega)$ (see Eq.~(1)) while that for WLI is determined by
$I(\Omega)$. For Gaussian spectra the effective bandwidth for CPI is
$\sqrt{2}$ narrower than that for the WLI, accounting for the
resolution advantage of Q-OCT.

Is this a fundamental feature of entanglement?  Surprisingly no. The
same difference in resolution could be achieved using purely
classical correlations. In CPI, time-correlations, but not intensity
correlations are created between anticorrelated frequencies.  In the
same setup, one replaces the chirped pulses with two CW lasers tuned
to frequencies $\omega_0 + \Omega$ and $\omega_0 + \Omega'$. During
the integration time of the detection the laser frequencies are
swept in an anticorrelated way according to the distribution
$P(\Omega,\Omega')=G(\Omega)\delta(\Omega+\Omega')$. In this case,
the effective spectrum in Eq.~(1) is $G(\Omega)$, while the WLI
bandwidth is determined by the marginal $P(\Omega)=\int d\Omega'
P(\Omega,\Omega')=G(\Omega)$. The effective bandwidths are identical
in both cases and thus this classical scheme achieves the same
factor of $2$ higher resolution as Q-OCT.  In practice, however,
chirped pulses offer dramatically higher nonlinear conversion
efficiency compared to CW lasers far outweighing this rather small
difference.

\section{Conclusion}
We have shown that chirped-pulse interferometry accrues the benefits
of quantum-OCT, but with a dramatic increase in signal, direct
optical detection, and a straightforward means of identifying
artifacts.  We have experimentally demonstrated improved resolution
in CPI over WLI with and without mismatched dispersion by up to a
factor of 2.65. Increasing the bandwidth of the light source is
required for the resolution in CPI-based OCT to compete with
established techniques.  Broadband (148nm) three-wave mixing of anticorrelated frequencies has been demonstrated in 1.5mm-thick nonlinear materials \cite{carrasco06}.  For safe \emph{in vivo} imaging, the optical
power can be attenuated before the sample without loss of signal
visibility \cite{kaltenbaek08}.  We have demonstrated SFG detection of reflected signals as low as 5mW \cite{kaltenbaek08}. Based on the results of \cite{peer07}, efficient SFG should be measureable with much lower reflected powers.
Incorporating better nonlinear
materials, such as PPKTP, and more sensitive detectors will further allow operation at lower power
levels. CPI achieves the benefits of quantum interferometry at
macroscopic power levels and represents a powerful new technique for
optical imaging. More generally, we have clarified the role of
entanglement versus correlation in axial imaging.

\section*{Acknowledgements}
We acknowledge financial support from NSERC and CFI.  We thank G.
Weihs and C. Couteau for the use of their spectrometer and technical
assistance.  We thank D. Biggerstaff, K. Bizheva, D. Strickland, and G. Weihs
for valuable discussions.  J.L. acknowledges financial support from the Bell Family Fund and
R.K. acknowledges financial support from IQC and ORDCF.


\begin{thebibliography}{99}
\bibitem{borosellmeier} Sellmeier coefficients, representative for Code 0211 microsheet glass,
were provided by Corning Inc.
\bibitem{widthfootnote} If the bandwidth in HOM interference is determined by a pair of
Gaussian bandpass filters in front of the detectors, rather than by the source, then the HOM
dip is narrower than the WLI by only a factor of $\sqrt{2}$.
\end{thebibliography}
\end{document}